# Novel High-Mobility Field-Effect Transistors Based on Transition Metal Dichalcogenides

V. Podzorov [*] and M. E. Gershenson
*Department of Physics and Astronomy, Rutgers University, Piscataway, New Jersey 08854*

Ch. Kloc, R. Zeis, and E. Bucher
*Bell Laboratories, Lucent Technologies, 600 Mountain Ave., Murray Hill, New Jersey 07974*

We report on fabrication of novel field-effect transistors (FETs) based on transition metal dichalcogenides. The unique structure of single crystals of these layered inorganic semiconductors enables fabrication of FETs with intrinsically low field-effect threshold and high charge carrier mobility, comparable to that in the best single-crystal Si FETs (up to 500 cm$^2$/Vs for the *p*-type conductivity in the WSe$_2$-based FETs at room temperature). These novel FETs demonstrate ambipolar operation. Owing to mechanical flexibility, they hold potential for applications in "flexible" electronics.

In modern electronics, the requirements to field-effect transistors are stringent and often contradictory: e.g., for many applications, a combination of high charge carrier mobility ($\mu$) and mechanical flexibility is desirable. Neither of the developed FETs satisfies these requirements. For example, the process of fabrication of silicon FETs with a relatively high $\mu \leq 500$ cm$^2$/Vs [1,2] is incompatible with flexible substrates. The organic-based FETs, that provide basis for flexible electronics [3,4,5], are notoriously known for their low $\mu$. Although several novel nano-structured materials have been recently considered for FETs [6,7,8], the demand for high-mobility flexible devices has not been satisfied.

In this paper, we report on the novel class of field-effect transistors based on transition metal dichalcogenides (TMDs). These materials with a layered crystal structure are uniquely positioned for the field-effect applications. On the one hand, similarly to other inorganic semiconductors, they demonstrate a high mobility of charge carriers. The high mobility, comparable to that in the best Si MOSFETs, is due to the strong covalent bonding of atoms within the layers. On the other hand, similarly to organic semiconductors, the interlayer bonding in TMD crystals is of a weak van der Waals type, which results in an intrinsically low density of surface traps, and, thus, in a low field-effect threshold. The TMD-based transistors can operate in both the *electron*- and *hole*-accumulation modes, depending on the polarity of the gate voltage - this ambipolar operation is rarely observed in high-mobility FETs. Finally, these FETs survive bending, owing to the mechanical flexibility of TMD crystals and the *parylene* gate dielectric [9].

The transition metal dichalcogenides belong to the class of layered inorganic semiconductors with a chemical formula *MX*$_2$, where *M* stands for a transition metal and *X* - for Se, S or Te [10,11]. Single crystals of TMDs are formed by stacks of *X-M-X* layers. Atoms within each layer are held together by strong covalent-ionic mixed bonds, whereas the layers are weakly bonded to each other by van der Waals forces. The bonding anisotropy defines the unique morphology of these crystals as thin, flexible and easy-to-cleave platelets with atomically smooth (*a,b*)-facets. The electronic properties of TMDs vary from semiconducting (e.g., WSe$_2$) to superconducting (e.g., NbSe$_2$) [10,11]. The semiconducting TMDs are considered to be promising materials for solar cells, photoelectrochemical cells and *p-n*-junctions [12,13,14].
We have successfully used several TMD materials for FET fabrication. However, not all of them demonstrated the same combination of characteristics: e.g., we observed only unipolar conductivity in devices based on MoSe$_2$ (*p*-type), SnS$_2$ and HfS$_2$ (*n*-type). Below we focus on characteristics of the *tungsten diselenide* FETs with the highest (in this group of materials) mobility. The WSe$_2$ single crystals with the basal plane dimensions up to 10 mm have been grown in closed ampoules using the vapor phase transport method with Se excess [13]. The source and drain contacts have been deposited on the atomically-flat (*a,b*)-facet of WSe$_2$ single crystals either by thermal evaporation of 300 Å-thick silver film in high vacuum (~ 5x10$^{-7}$ Torr), or by using a conductive silver paint. The thermally evaporated gate electrode was electrically isolated from the crystal and contacts with a 1-µm-thick parylene film; this process is similar to the fabrication of organic single-crystal FETs described by us in Refs. [9,15]. The typical channel dimensions were *L* x *W* = 1x1 mm$^2$. We have also studied the 4-probe FETs, where two additional voltage probes were located in the channel between the source and drain contacts (see the inset in Fig. 5). Measurements of the latter devices by the 4-probe technique allowed extracting the intrinsic field-effect mobility of the charge carriers, not limited by the contact resistance. In all our measurements, $V_S$ and $V_g$ were applied with respect to the grounded drain contact.

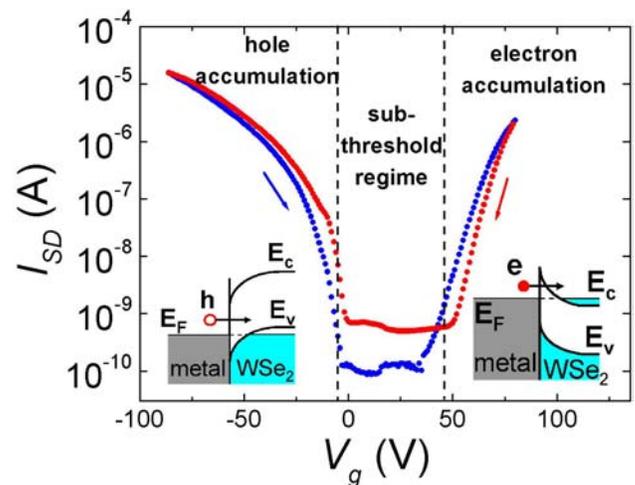

**FIG. 1.** The trans-conductance characteristics of the WSe$_2$ FET measured at $T = 60$ K. The observed hysteresis corresponds to sweeping $V_g$ in the opposite directions shown by arrows ($V_S = +10$ V). The insets illustrate bending of the valence (E$_V$) and conduction (E$_C$) bands of the semiconductor at the interface with the injecting contact; E$_F$ is the Fermi level in this contact.

The source-drain current ($I_{SD}$) as a function of gate voltage ($V_g$), measured at a fixed source-drain voltage ($V_S$) at 60 K, is shown in Fig. 1. At low temperatures, where the bulk

---

conductivity of the studied WSe$_2$ crystals is small, the on/off ratio exceeds $10^4$. The gate-induced charge injection from contacts is manifested by the sharp increase of $I_{SD}$ at both negative (the *p*-type conductivity) and positive (the *n*-type conductivity) $V_g$. A non-zero threshold voltage, $V_g^{th}$, indicates that, despite the high quality of the basal surface of WSe$_2$, a small density of charge traps ($< 5 \times 10^{10}$ cm$^{-2}$) is still present at the semiconductor/dielectric interface. Residual conductivity in the sub-threshold regime is mostly due to the bulk conductivity of WSe$_2$ crystals, which is relatively large at high temperature because of unintentional bulk doping of the studied WSe$_2$ crystals (see Fig. 2).

The ambipolar operation is expected in the Schottky-type FETs, where the transistor action occurs by modulating the tunneling current through the Schottky barriers, formed at the metal-semiconductor contacts. The barrier thickness and the sign of injected charges are controlled by "bending" the electronic bands at the semiconductor surface by the gate voltage (as sketched in the insets to Fig. 1). Though several Schottky-type FETs are known, the reports of ambipolar operation in symmetric devices with a single conducting channel are rare. Examples of ambipolar devices include the amorphous silicon ($\alpha$-Si:H) FETs with rather low electron and hole mobilities (0.1 and $2 \times 10^{-3}$ cm$^2$/Vs, respectively) [16], and the FETs based on carbon nanotubes [17,18,19]. Ambipolar transport with very low mobilities has been also observed in organic FETs [20] and organic hetero-structure FETs with two different active layers [21]. In the latter case, however, the hole and electron currents are spatially separated. The mobilities observed in the reported ambipolar devices are typically very low, except for the carbon nanotube FETs, where the charge carrier transport is ballistic.

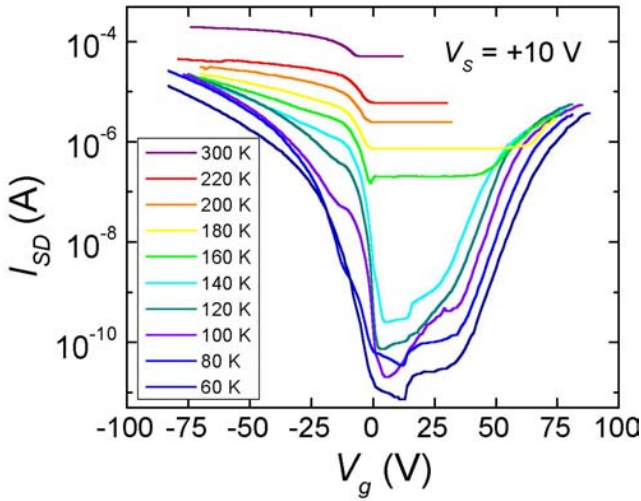

**FIG. 2.** The trans-conductance characteristics of WSe$_2$ FET measured at different temperatures, at a constant source-drain voltage of fixed polarity.

The asymmetry of the trans-conductance characteristics with respect to $V_g = 0$ in Fig. 1 is caused by substantially different thresholds for *n*- and *p*-type operation, as well as by the positive polarity of the source potential ($V_S = +10$ V). The hysteresis in $I_{SD}(V_g)$ has been observed at low temperatures ($T \leq 100$ K). Similarly to the case of $\alpha$-Si:H FETs [22], this hysteresis reflects slow "re-charging" of the surface traps at the time scale comparable to the measurement time (in our measurements, the $V_g$ sweep rate was 5 V/min).

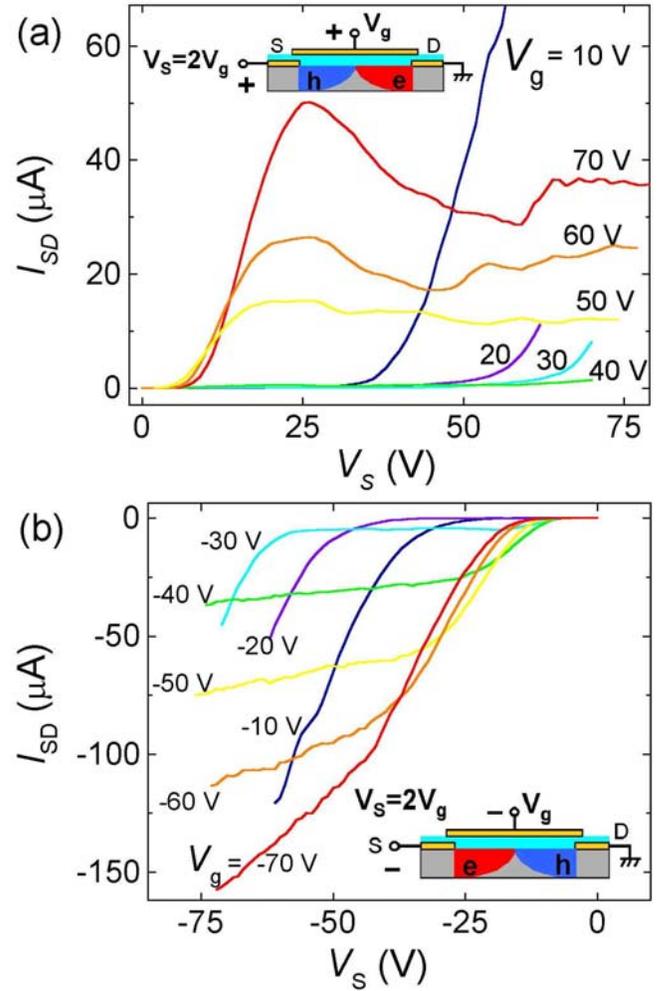

**FIG. 3 a, b.** $I_{SD}(V_S)$ characteristics of the WSe$_2$ FET for several positive (a) and negative (b) values of $V_g$. Two types of the I-V curves are clearly seen: non-saturating at small $V_g$ ($< 40$ V) and large $V_S$, and saturating at larger $V_g$. The former regime corresponds to the formation of *p*-type channel, whereas in the latter regime, both *n*- and *p*-type regions coexist in the channel (see the text). The insets illustrate simultaneous formation of the *n*- and *p*-type conduction regions at $V_S = 2V_g$.

In the ambipolar devices, the electrons and holes can be injected simultaneously into the opposite ends of the channel. This interesting situation can be realized if the transverse electric field has opposite signs near the source and drain contacts (as shown in the insets to Fig. 3 a,b). Figure 3 illustrate this regime, it shows the $I_{SD}(V_S)$ dependences measured at 120 K for several positive (negative) values of $V_g$ and $V_S$. In Fig. 3 a, at relatively small $V_g < 40$ V and large $V_S > V_g$, the gate is *negatively* biased with respect to the source, and the *p*-type accumulation layer is formed near the source contact. The injection of holes from the source is manifested by a steep increase of the current with $V_S$. The electron injection from the drain is suppressed in this biasing regime because of a relatively large threshold $V_g^{th}(n)$ for the *n*-type operation. When $V_g$ exceeds $V_g^{th}(n)$, the shape of $I_{SD}(V_S)$ characteristics in Fig. 3 changes dramatically. At large gate voltages ($V_g > 50$ V), initial increase of $I_{SD}$ at small $V_S < 25$ V corresponds to injection of electrons from the drain and formation of the *n*-type channel. With further increase of $V_S$, the *p*-type accumulation layer is formed near the source and the carriers of opposite signs are simultaneously injected from the source and drain contacts. This regime is especially interesting because holes and electrons, moving within the same channel in the opposite directions, can recombine. Since $WSe_2$ can be considered as an "almost" direct band-gap semiconductor [13], the electron-hole recombination might result in light emission. This very interesting possibility of creating a light-emitting transistor requires further studies. It is worth mentioning that light emission has been observed in ambipolar carbon nanotube FETs [18].

Figure 3 b shows the $I_{SD}(V_S)$ characteristics for several negative values of $V_g$. Again, at small $|V_g| < 30$ V and large negative $V_S$, the electrons are injected from the source contact, and the n-channel formation is manifested by a rapid increase of $|I_{SD}|$. However, at $|V_g| > 40$ V, the conduction is dominated by the drain-injected holes. With increasing $|V_S|$ in this regime, $I_{SD}$ exhibits the expected saturation due to the pinch-off of the conducting channel.

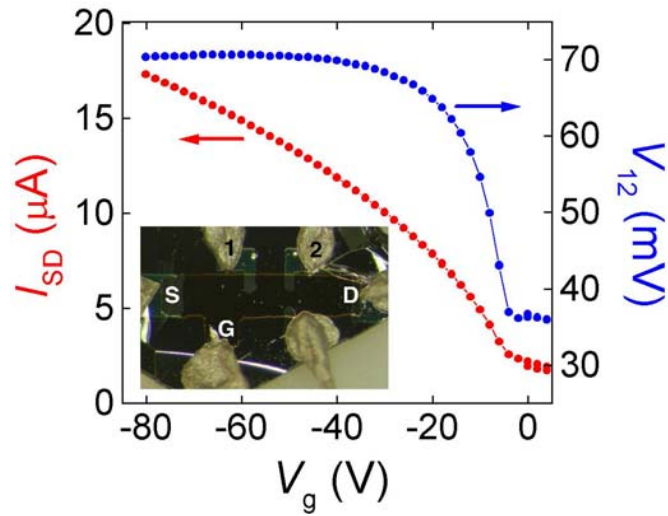

**FIG. 4.** Characteristics of 4-probe $WSe_2$ FET. The dependences $I_{SD}(V_g)$ and $V_{12}(V_g)$, measured for a 4-probe $WSe_2$ FET at a fixed source-drain voltage ($V_S = 1$ V) at room temperature. The inset is a photograph of one of the 4-probe devices with the source (S), drain (D), gate (G) electrodes and two voltage probes (1 and 2).

For extracting the intrinsic mobility of charge carriers in $WSe_2$ FETs, i.e. the mobility that is not limited by the contact resistance, we have performed the 4-probe measurements. The $V_g$ –dependences of $I_{SD}$ and the potential difference between the voltage probes ($V_{12}$) for such a gated 4-probe device are shown in Fig. 4. These dependences have been measured at room temperature at fixed $V_S = 1$ V. With increasing $|V_g|$, the Schottky contact resistance decreases sharply, and the voltage drop across the middle section of the channel $V_{12}$ increases. For sufficiently large $|V_g|$, the dependence $I_{SD}(V_g)$ becomes quasi-linear, which corresponds to a $V_g$ – independent mobility of the charge carriers. In this regime, the intrinsic mobility ~ 500 cm$^2$/Vs was obtained using the equation for the 4-probe device $\mu = [D/(WC_i)]*[\mathbf{d}((I_{SD}-I_0)/V_{12})/dV_g]$, where $C_i = 2 \pm 0.2$ nF/cm$^2$ is the capacitance between the gate and the channel, $D$ is the distance between the voltage probes 1 and 2, and $I_0$ is the current in the subthreshold regime [9,15]. Note that the "apparent" mobility, estimated from Fig. 4 using the conventional 2-probe expression, $\mu = [L/(WC_iV_S)]*[\mathbf{d}I_{SD}/\mathbf{d}V_g]$, is lower (~ 100 cm$^2$/Vs) because of the contact resistance.

Interestingly, the TMD-based FETs are flexible owing to the outstanding flexibility of both the layered semiconductor and the parylene gate insulator. We observed that bending of the $WSe_2$ FETs did not significantly affect their performance in the p-channel mode. The n-type conductivity diminishes after a few bending cycles, this might be caused by strain-induced increase of the Schottky barrier for the electron injection, or by modification of the band gap of the semiconductor with strain. Similar behavior has been observed in the carbon-nanotube FETs [19].

In conclusion, a novel class of high-mobility flexible inorganic FETs based on transition metal dichalcogenides has been demonstrated. These layered materials combine the advantages of organic and inorganic semiconductors, providing surfaces with an intrinsically low density of traps and high carrier mobility. The mobility of the field-induced holes in $WSe_2$-based devices is ~500 cm$^2$/Vs at room temperature, which is comparable to the mobility of electrons in the best (non-flexible) Si MOSFETs and exceeds $\mu$ in organic TFTs by ~ 3 orders of magnitude. Provided that the doping level in $WSe_2$ can be reduced by further optimization of the crystal growth, a low room-temperature bulk conductivity and, hence, a high on/off ratio can be realized. In contrast to the Si devices, the $WSe_2$ FETs can operate as ambipolar devices. These promising electrical characteristics, in combination with mechanical flexibility, make the field effect transistors based on transition metal dichalcogenides very attractive for "flexible" electronics.

This work at Rutgers University has been supported by the NSF grant DMR-0077825.